\documentclass[aps,prc,preprint,groupedaddress,showpacs]{revtex4}

\usepackage{graphicx}

\begin{document}

\title{Constraining the density dependence of the symmetry energy in the nuclear equation
of state using heavy ion beams}
\author{D.V. Shetty, S.J. Yennello, and G.A. Souliotis}
\affiliation{Cyclotron Institute, Texas A$\&$M University, College Station, TX 77843, USA}
\date{\today}

\begin{abstract}
The density dependence of the symmetry energy in the equation of state of asymmetric nuclear 
matter (N/Z $>$ 1) is important for understanding the structure of systems  as diverse as the atomic 
nuclei and neutron stars.  Due to a proper lack of understanding of the basic nucleon-nucleon 
interaction for matters that are highly asymmetric and at non-normal nuclear density, this very 
important quantity has remained largely unconstrained. Recent studies using beams from the 
Cyclotron Institute of Texas A$\&$M University, constraining the density dependence of the symmetry 
energy, is presented. A dependence of the form E$_{sym}(\rho)$ = C($\rho/\rho_{o})^{\gamma}$, where 
C = 31.6 MeV and $\gamma$ = 0.69, is obtained from the dynamical and statistical model analysis. 
Their implications to both astrophysical and nuclear physics studies are discussed.
\end{abstract}


\maketitle

\section{Introduction}

The fundamental goal of nuclear physics is to understand the basic building blocks of nature - neutrons 
and protons - and the force of interaction that binds them together into nuclear matter. Studying the  
nature of matter and the strength of nuclear  interaction is key to understanding some of the fundamental 
problems such as, how elements are formed ? How stars explode into supernova ? What kind of matter exists 
inside a neutron star ? How neutrons are compressed inside the neutron star to density trillions of times 
greater than on earth ? What determines their density-pressure relation, the so-called equation of state ? 
\par
Ordinary matter and their equation of state (the real gas equation) are well understood in terms 
of Van der Waal's inter-molecular force of attraction. The nuclear matter equation of state on 
the other hand, is not so well understood due to a proper lack of understanding of the basic 
nucleon-nucleon interaction over a wide range of density, temperature and isospin 
(neutron-to-proton ratio). Until now our understanding of the nucleon-nucleon interaction has 
come from studying nuclear matter  that are  symmetric in isospin ({\it {i.e.}}, neutron-to-proton 
ratio equal one) and those found near normal nuclear matter density ($\rho_{o}$ = 0.16 $fm^{-3}$). 
It is not known how far these understanding remains valid as one go away from the normal nuclear 
density and symmetric nuclear matter. Various interactions used in `` ab initio " microscopic 
calculations predict different forms of the nuclear equation of state above and below the normal nuclear 
matter density, and away from the symmetric nuclear matter \cite{FUC}.  As a result, the symmetry 
energy, which is the difference in energy between the pure neutron matter and the symmetric 
nuclear matter, show very different behavior above and below normal nuclear density. Fig. 1 shows the 
the density dependence of the symmetry energy as predicted by various microscopic calculations. One 
observes from the figure that in general two different forms are predicted. One, where the 
symmetry energy increases monotonically with increasing density (`` stiff " dependence) and the other, 
where the symmetry energy increases initially up to normal nuclear density and then decreases at 
higher densities (`` soft " dependence). Constraining the form of the density dependence of the 
symmetry energy is important not only for a better understanding of the nucleon-nucleon interaction, 
and hence their extrapolation to the  structure of neutron-rich nuclei, but also to determine the 
structure of compact stellar objects such as the neutron stars. For example, a `` stiff " density 
dependence of the symmetry energy is predicted to lead to a large neutron skin thickness compared to 
a `` soft " dependence \cite{OYA98}. Similarly, a `` stiff " dependence of the symmetry energy can result 
in rapid cooling of a neutron star, and a larger neutron star radius, compared to a soft density 
dependence \cite{LAT94}.
\par
Experimentally, one possible  means of studying the  nuclear equation  of state  at sub-normal  nuclear 
density and  high excitation energy is through  the intermediate  energy heavy ion collision reactions. 
In these reactions, an excited nucleus (the composite of the projectile and the target nucleus) expands 
to a sub-nuclear density  and disintegrates into  various light and heavy fragments in a process called 
multifragmentation. By studying the isotopic yield  distribution  of these  fragments one  can  extract 
important information about the symmetry energy and their density dependence. 

\begin{figure*}
\resizebox{0.8\textwidth}{!}{%
\includegraphics{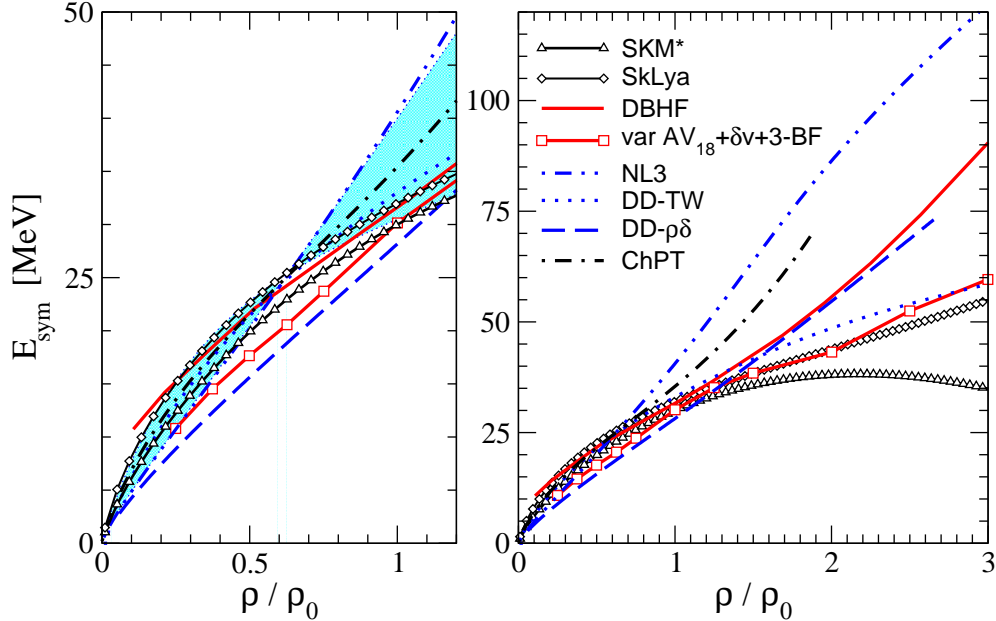}
}
\caption{\label{figure1} Symmetry energy as a function of density as predicted by different theoretical
models. The left panel shows the low density region and the right panel shows the high density range. The figure
is taken from Ref. \cite{FUC}}
\end{figure*} 

\section{Symmetry energy and the isotopic yield distribution}

It has been shown from experimental observations that the ratio of the fragment isotopic yields in two 
different reactions, 1 and 2, R$_{21}$($N$,$Z$) = Y$_{2}$($N$,$Z$)/Y$_{1}$($N$,$Z$), obey an exponential 
dependence on the neutron number ($N$) and the proton number ($Z$) of the isotopes; an observation known 
as isoscaling \cite{BOT02, TSA01}. The dependence is characterized by the relation,

\begin{equation}
       R_{21}(N,Z) = Ce^{(\alpha N + \beta Z)}
\end{equation}

Where, Y$_{2}$ and Y$_{1}$ are the yields from the neutron-rich and neutron-deficient systems, respectively.
$C$ is an overall normalization factor, and $\alpha$ and $\beta$ are the parameters characterizing the 
isoscaling behavior. The observation is also theoretically predicted by the statistical and the dynamical 
multifragmentation models \cite{BOT02,TSA01,ONO03}. In these models, the difference in the chemical potential of systems with 
different $N/Z$ is directly related to the scaling parameter $\alpha$, which is further related to the 
symmetry energy through a relation, 

\begin{equation}
           \alpha = \frac{4C_{sym}}{T} \bigg (\frac{Z_{1}^{2}}{A_{1}^{2}} - \frac{Z_{2}^{2}}{A_{2}^{2}}\bigg )
\end{equation}

where, $Z_{1}$, $A_{1}$ and $Z_{2}$, $A_{2}$ are the charge and the mass numbers of the fragmenting systems,
$T$ is the temperature of the system and $C_{sym}$, is the symmetry energy.

\section{Experiment}

The isotopic yield distributions for the present study were obtained by carrying out measurements at the 
Cyclotron Institute of Texas A$\&$M University (TAMU), using the K500 Superconducting Cyclotron, and the 
National Superconducting Cyclotron Laboratory (NSCL) at Michigan State University (MSU). Reactions such as, 
$^{40}$Ar, $^{40}$Ca + $^{58}$Fe, $^{58}$Ni and $^{58}$Ni, $^{58}$Fe + $^{58}$Ni, $^{58}$Fe,  at beam energies 
from 25 MeV/nucleon to 53 MeV/nucleon, were studied to measure the fragment yield distribution. Details of 
the measurement can be found in \cite{IGL06}, and references therein.

\section{Results and Discussions}

\begin{figure*}
\resizebox{0.7\textwidth}{!}{%
\includegraphics{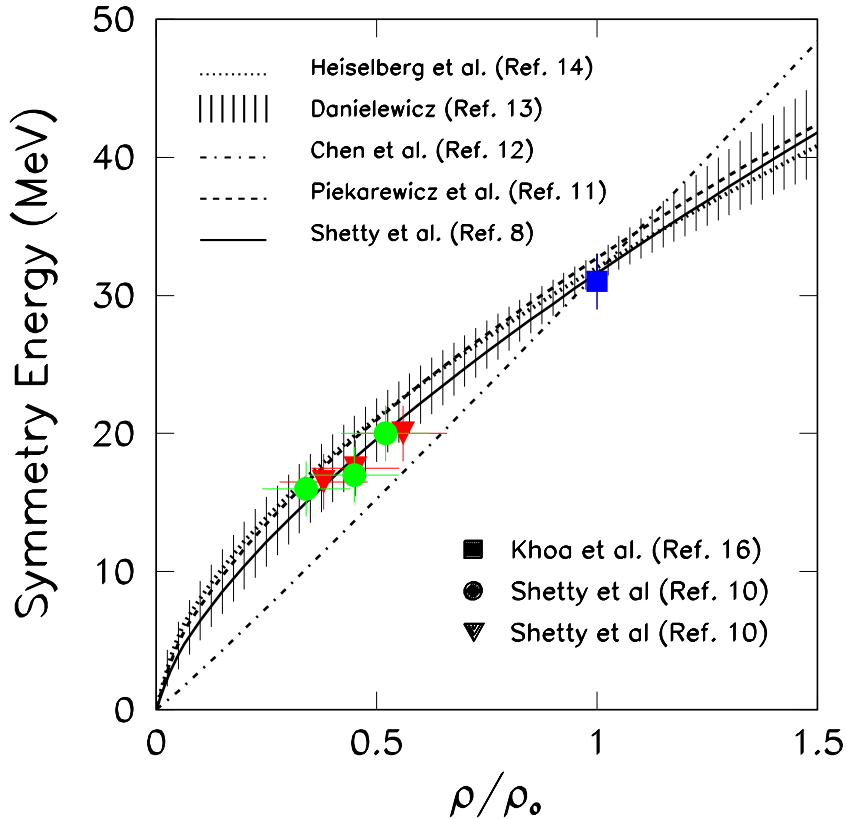}
}
\caption{\label{figure2} Comparison of the density dependence of the symmetry energy obtained from various 
different studies. See text for details and table I.}
\end{figure*} 

Fig. 2 shows the form of the density dependence of the symmetry energy obtained by comparing the 
experimentally measured isoscaling parameter with those obtained from the statistical and the dynamical 
multifragmentation models. Also shown in the figure are comparisons with several other recent independent 
studies. The solid curve corresponds to the one obtained from Gogny-AS interaction in dynamical AMD model 
that explains the present results, assuming the sequential decay effect to be small \cite{SHE04, SHE05}. The triangle and 
the circle symbols also correspond to the present measurements obtained by comparing with the statistical 
multifragmentation model (see Ref. \cite{SHE06} for more details). The dashed curve correspond to those obtained recently from an accurately 
calibrated relativistic mean field interaction, used for describing the Giant Monopole Resonance (GMR) in
$^{90}$Zr and $^{208}$Pb, and the IVGDR in $^{208}$Pb by Piekarewicz {\it {et al}}. \cite{PIE05}. The dot-dashed 
curve correspond to the one used for explaining the isospin diffusion data of NSCL-MSU by Chen {\it {et al.}} 
\cite{CHE05}. This dependence has now been modified to include the isospin dependence of the in-medium nucleon-nucleon 
cross-section, and in good agreement with the present study. The shaded region in the figure corresponds to 
those obtained by constraining the binding energy, neutron skin thickness and isospin analogue state in 
finite nuclei using the mass formula of Danielewicz \cite{DAN04}. The dotted curve correspond to the parameterization 
adopted by Heiselberg {\it {et al.}} \cite{HEI00} in their studies on neutron stars, and obtained by fitting the 
predictions of the variational calculations of Akmal {\it {et al.}} \cite{AKM98}. Finally, the solid square point in 
the figure correspond to the value of symmetry energy obtained by fitting the experimental differential 
cross-section data in a charge exchange reaction using the isospin dependent optical potential by Khoa {\it {et
al.}} \cite{KHO05}. The parameterized forms of the density dependence of the symmetry energy obtained from all these 
studies are as shown in table I. The  close agreement between various independent studies show that a 
constraint on the density dependence of the symmetry energy, given as E$_{sym}(\rho)$ = C($\rho/\rho_{o})^{\gamma}$, 
where C = 31.6 MeV and $\gamma$ = 0.69 can be obtained.

\section{Conclusions}

A number of reactions were studied at TAMU to investigate the density dependence of the symmetry 
energy in the equation of state of asymmetric nuclear matter. The results were analyzed within the
framework of dynamical and statistical model calculations. It is observed that a dependence of the form
E$_{sym}(\rho)$ = 31.6 ($\rho/\rho_{o})^{0.69}$ MeV, agrees better with the experimental data, indicating 
that a `` stiff " form of the nucleon-nucleon interaction provides a better understanding of the nuclear 
matter EOS. The observed constrain leads to nuclear matter compressibility $K$ $\sim$ 230 MeV, and a neutron 
skin thickness $R_{n}$ - $R_{p}$ $\sim$ 0.21 fm, for $^{208}$Pb nuclei. It also predicts a neutron 
star mass of 1.72 solar mass and a  radius, $R$ = 11 - 13 km for the `` canonical " neutron star. 
Furthermore, it predicts a direct URCA cooling for neutron stars above 1.4 times the solar mass. These 
results have important implications for nuclear astrophysics and future experiments probing the properties 
of nuclei using beams of neutron-rich nuclei.

\begin{table*}
\caption{\label{tab:table1}Parametrized form of the density dependence of the symmetry energy obtained
from various independent studies.}
\begin{ruledtabular}
\begin{tabular}{cccccccc}
   Reference                                     &  &   Parameterization                         &  Studies                             \\
\hline
Heiselberg {\it {et al.}}  \cite{HEI00}          &  &  32.0($\rho$/$\rho_{o}$)$^{0.60}$          &  Variational calculation             \\
Danielewicz {\it {et al.}} \cite{DAN04}          &  &  31(33)($\rho$/$\rho_{o}$)$^{0.55(0.79)}$  &  BE, skin, isospin analog states     \\
Chen {\it {et al.}}        \cite{CHE05}          &  &  31.6($\rho$/$\rho_{o}$)$^{1.05}$          &  Isospin difussion                   \\  
Piekarewicz {\it {et al.}} \cite{PIE05}          &  &  32.7($\rho$/$\rho_{o}$)$^{0.64}$          &  Giant resonances                    \\ 
Shetty {\it {et al.}}      \cite{SHE04, SHE06}   &  &  31.6($\rho$/$\rho_{o}$)$^{0.69}$          &  Isotopic distribution            
\end{tabular}
\end{ruledtabular}
\end{table*}

\section{Acknowledgements}
This work was supported in part by the Robert A. Welch Foundation through grant No. A-1266, and the Department of 
Energy through grant No. DE-FG03-93ER40773.


\begin{thebibliography}{}
\bibitem{FUC} C. Fuchs and H.H. Wolter, nucl-th/0511070 (2005), Euro. Phys. J. A (In press).   
\bibitem{OYA98} K. Oyamatsu {\it {et al.}}, Nucl. Phys. A{\bf 634}, (1998) 3.  
\bibitem{LAT94} J. Lattimer {\it {et al.}}, Astr. Phys. Jour. {\bf 425}, (1994) 802.  
\bibitem{BOT02} A.S. Botvina {\it {et al.}}, Phys. Rev. C {\bf 65}, (2002) 044610. 
\bibitem{TSA01} M.B. Tsang {\it {et al.}}, Phys. Rev. C {\bf 64}, (2001) 054615.    
\bibitem{ONO03} A. Ono {\it {et al.}}, Phys. Rev. C {\bf 68}, (2003) 051601.        
\bibitem{IGL06} J. Iglio {\it {et al.}}, Phys. Rev. C {\bf 74}, 024605 (2006).     
\bibitem{SHE04} D.V. Shetty {\it {et al.}}, Phys. Rev. C {\bf 70}, (2004) 011601.    
\bibitem{SHE05} D.V. Shetty {\it {et al.}}, nucl-ex/0512011 (2005).            
\bibitem{SHE06} D.V. Shetty {\it {et al.}}, nucl-ex/0606032 (2006).            
\bibitem{PIE05} J. Piekarewicz (Private Communication) (2005).        
\bibitem{CHE05} L.W. Chen {\it {et al.}}, Phys. Rev. Lett. {\bf 94}, (2005) 032701.   
\bibitem{DAN04} P. Danielewicz, nucl-th/0411115 (2004).                  
\bibitem{HEI00} H. Heiselberg and M. Hjorth-Jensen, Phys. Rep. {\bf 328}, (2000) 237. 
\bibitem{AKM98} A. Akmal {\it {et al.}}, Phys. Rev. C {\bf 58}, (1998) 1804.  
\bibitem{KHO05} D.T. Khoa {\it {et al.}}, Phys. Rev. C {\bf 71}, (2005) 044601.  
  
\end{thebibliography}
\end{document}